\documentstyle{cupconf}
\input psfig.sty

\ifoldfss
\else
  \ifnfssone
    \newmathalphabet{\mathit}
      \addtoversion{normal}{\mathit}{cmr}{m}{it}
      \addtoversion{bold}{\mathit}{cmr}{bx}{it}
    \newmathalphabet{\mathcal}
      \addtoversion{normal}{\mathcal}{cmsy}{m}{n}
    \else
    \ifnfsstwo
    \fi
  \fi
\fi

\def\hexnumber#1{\ifcase#1 0\or1\or2\or3\or4\or5\or6\or7\or8\or9\or
 A\or B\or C\or D\or E\or F\fi }

\makeatletter
\ifx\CUP@mtlplain@loaded\undefined
\else
\fi
\makeatother
 \makeatletter
 \ifx\CUP@mtlplain@loaded\undefined
   \font\tenbmi=cmmib10 at 10pt
   \font\sevenbmi=cmmib10 at 7pt
   \font\fivebmi=cmmib10 at 5pt

   \newfam\bmifam
   \textfont\bmifam=\tenbmi
   \scriptfont\bmifam=\sevenbmi
   \scriptscriptfont\bmifam=\fivebmi
   
 \fi
 \makeatother
\ifnfsstwo

\fi
\ifnfssone

\fi
\ifoldfss

\fi

\mathchardef\varLambda="0103

\makeatletter
\ifx\CUP@mtlplain@loaded\undefined
\else
\fi
\makeatother
\makeatletter
\ifx\CUP@mtlplain@loaded\undefined
  \font\tenbms=cmbsy10
  \font\sevenbms=cmbsy10 at 7pt
  \font\fivebms=cmbsy10 at 5pt
  \newfam\bmsfam
  \textfont\bmsfam=\tenbms
  \scriptfont\bmsfam=\sevenbms
  \scriptscriptfont\bmsfam=\fivebms

  \edef\bsy@{\hexnumber\bmsfam}
  \mathchardef\bnabla="0\bsy@72
\fi
\makeatother
\title[]{Radio cores in blazars}

\author[T. Venturi {\it et al.\/}]%
{T. V\ls E\ls N\ls T\ls U\ls R\ls 
I$^1$, \ns D.\ls D\ls A\ls L\ls L\ls A\ls C\ls A\ls S\ls A$^2$\ns\\
S.\ls T\ls O\ls R\ls R\ls I$^2$ \and \ns F.\ls M\ls A\ls N\ls
T\ls O\ls V\ls A\ls N\ls I$^1$}

\affiliation{$^1$Istituto di Radioastronomia, CNR, Via Gobetti
101, 40129 Bologna, Italy\\
$^2$Astronomy Dept., University of Bologna,
Via Ranzani 1, 40127 Bologna, Italy}

\begin{document}
\ifnfssone
\else
  \ifnfsstwo
  \else
    \ifoldfss
      \let\mathcal\cal
      \let\mathrm\rm
      \let\mathsf\sf
    \fi
  \fi
\fi

\maketitle

\begin{abstract}
We present results and first epoch parsec-scale images from a monitoring 
program  on five $\gamma$--ray loud blazars selected on the basis
of their total flux density variability. We observed
0048$-$097, 0235+164, 0954+658, 1510$-$089 and 1749+096
with the Very Long Baseline Array simultaneously at 8.4 GHz and
22 GHz. Comparison between our high quality images and 
previous results from the literature suggest that 0954+658 and
1510$-$089 are superluminal radio sources, with a Lorentz
factor $\gamma$ of the order of $\sim 2 - 3$.
\end{abstract}

\firstsection 
\section{Introduction}
Flux density variability at high frequencies in compact
extragalactic radio sources is usually intrinsic and contains 
important pieces of information on the size of the emitting
region, the orientation of radio plasma to the line of
sight and its intrinsic plasma speed. Furthermore, the delay of
radio flares going from high to low radio frequencies, coupled
with the knowledge of other parameters, such as, for example,
the optical depth, is crucial to locate its position within
the radio source. Variable radio sources are the best candidates
for the detection of structural changes and proper motion, and to 
derive estimates on their intrinsic (local) Lorentz factor.

In this paper we present VLBA radio images and preliminary results
for five $\gamma$--ray loud blazars, i.e.
0048$-$097, 0235+164, 0954+658, 1510$-$089 and 1749+096.
These sources were selected from a sample of variable $\gamma$--ray loud 
blazars monitored at 8.4 GHz and 5 GHz with the VLBI antenna
located in Medicina (Bologna, Italy) (Venturi et al. 1999).
1510$-$089 is a high
polarisation quasar, while all other sources are BL-Lacs.
They all showed strong variability since 1996 in the form of {\it (a)} 
major radio flares (0235+164); {\it (b)} periodic variations (0048$-$097);
{\it (c)} complex light curves (the remaining sources).

\section{Observations and Images}

We observed these sources with the Very Long Baseline Array
(VLBA) in snapshot, dual polarisation mode at 8.4 GHz and at 22 GHz 
in January 1999. The total integration time on source was $\sim 2$ hours
each frequency. The angular resolution of our observations is
$\le$ 1 mas at both frequencies and the rms in the final images
ranges from 0.07 mJ/beam to 1.8 mJy/beam, depending on the 
frequency and peak in the image.

0048$-$097 and 0235+164 are unresolved at both frequencies and have 
a flat spectrum, i.e. $\alpha \sim 0.3$, in the range of our
observations. The three remaining sources exhibit a core-jet
morphology, suitable for the detection of proper motion.

\begin{figure} 
   \vspace{3pc}
\psfig{file=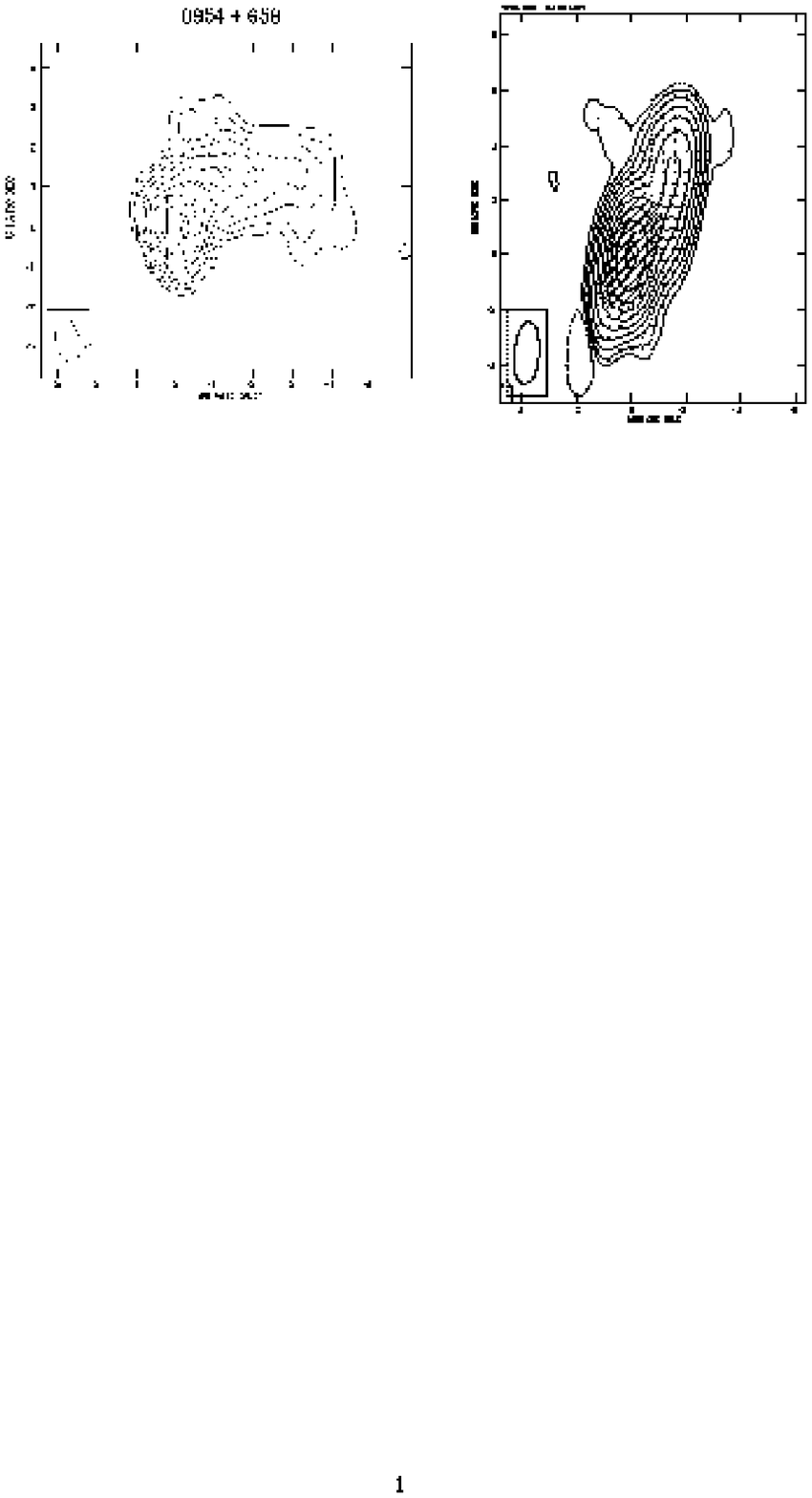}
  \caption{Left - 8.4 GHz image of 0954+658.
The restoring beam is 1.07$\times$0.65 mas, in p.a. $14^{\circ}$; 
the rms in the image is 0.07 mJy/b and the contours are 0.07 $\times~
-3$, 3, 6, 12, 24, 96, 192, 375 mJy/b. The polarisation vectors
(electric vector) is superposed. Right - 8.4 GHz image of 1510$-$089
with electric field vectors superposed. 
The restoring beam is 2.29$\times$0.88 mas, in p.a. $-4.6^{\circ}$;
the rms in the image is 0.26 mJy/b and the contours are 0.26 $\times~
-3$, 3, 6, 12, 24, 96, 192, 375 mJy/b.}
\end{figure} 

\section{Preliminary Results}

The two sources shown in Figure 1 are the best candidates for the
detection of superluminal motion. Second and third epoch monitoring
with VLBA are in the observing queue and they will allow
proper image comparison and analysis. At this stage we searched for
images at comparable frequency and resolution from the literature
(Gabuzda et al. 1994) and from the VLBA geodetic database 
(Fey et al. 1997). On the basis of our preliminary component 
identification we found that both sources have superluminal
components. The innermost component along the jet of 0954+658 has
a superluminal speed with $\beta_{app} \sim$ 1.3, leading to a
minimum value of the Lorentz factor $\gamma_{min} \sim 1.7$.
In the case of 1510$-$089 we identified two knots along the
jet, and we derived $\beta_{app} \sim 2.7$ and 1.6 
moving outwards along the jet. The inferred Lorentz factors are
$\sim$ 1.9 - 2.8. 

The images shown in Figure 1 show the orientation of the magnetic
field assuming that Faraday rotation is negligible.
In 0954+658 the field is parallel to the jet direction,
as expected for BL-Lac objects. The case of 1510$-$089 is
more complicated. Here the magnetic field rotates along the
jet and from the inner to the outer part. It is interesting
to note that this source is also characterised by a high
fractional polarisation, i.e. $\sim 7$\%.

\end{document}